\documentclass[11pt]{article}
\usepackage{amsmath,amssymb,color,epsfig,cite}
\usepackage{graphicx}
\usepackage{subfigure}
\usepackage{setspace}
\usepackage[arrow, matrix, curve]{xy}

\usepackage{amsfonts}

\makeatletter
\@addtoreset{equation}{section}
\makeatother

\DeclareTextFontCommand{\textwasy}{\wasyfamily}
\def \wasyfamily{\fontencoding{U}\fontfamily{wasy}\selectfont}
\def \thorn{{\wasyfamily\char105}}
\DeclareTextCommand{\dh}{OT1}{{\wasyfamily\char107}}
\newcommand{\tho}{{\textrm\thorn}}

\newcommand{\be}{\begin{equation}}
\newcommand{\ee}{\end{equation}}
\newcommand{\bea}{\setlength\arraycolsep{2pt} \begin{eqnarray}}
\newcommand{\eea}{\end{eqnarray}}

\newcommand{\ua}{\underline{a}}

\newcommand{\ba}{\boldsymbol{a}}
\newcommand{\bb}{\boldsymbol{b}}
\newcommand{\bc}{\boldsymbol{c}}

\newcommand{\bA}{\boldsymbol{A}}

\newcommand{\pa}{\phantom{a}}
\newcommand{\pb}{\phantom{b}}
\newcommand{\pc}{\phantom{c}}
\newcommand{\pd}{\phantom{d}}

\newcommand{\pA}{\phantom{A}}

\begin{document}

\vspace*{35pt}
\begin{center}
{\Large {\bf Penrose quasi-local energy and Kerr-Schild metrics}}

\vspace{15pt}
{\bf Mahdi Godazgar, Andr\'e Kaderli
}

\vspace{10pt}

{\it Institut f\"ur Theoretische Physik,\\
Eidgen\"ossische Technische Hochschule Z\"urich, \\
Wolfgang-Pauli-Strasse 27, 8093 Z\"urich, Switzerland.}
\medskip

{\it godazgar@phys.ethz.ch,\ \ kaderlia@student.ethz.ch}

 \vspace{20pt}

\vspace{20pt}

\underline{ABSTRACT}
\end{center}
 {\small \noindent Specialising to the case of Kerr-Schild spacetimes, which include the Kerr black hole and gravitational wave solutions, we propose a modification of the Penrose quasi-local energy.  The modification relies on the existence of a natural Minkowski background for Kerr-Schild spacetimes.  We find that the modified surface integral reduces to a volume integral of the Einstein tensor, which has been proposed previously as an appropriate definition for quasi-local energy for Kerr-Schild backgrounds.  Furthermore, in the special case that the Kerr-Schild null vector is normal to the surface of interest, we construct a 1-1 map between the 2-surface twistors in the Kerr-Schild background and Minkowski twistors projected onto the surface.}

\thispagestyle{empty}

\pagebreak



\section{Introduction}
A reasonable notion of gravitational energy~\footnote{It should be emphasised that amenability to relative ease of computation is an important aspect of reasonability.
} remains an unresolved issue in the context of general relativity more than a century after its discovery.~\footnote{See Ref. \cite{Szabados:2004vb} for a review of (quasi-local) energy-momentum constructions in GR.} The main difficulty is the lack of a background geometry; a difficulty shared with another important outstanding problem in general relativity: that of quantisation.  This means that the well-established Noether-Belinfante-Rosenfeld procedure for the energy-momentum density of matter fields cannot be generalised to include the gravitational field.  Indeed, due to the principle of equivalence any tensorial gravitational energy-momentum is expected to vanish locally. Different coordinate-dependent approaches have been suggested, such as energy-momentum pseudotensors and bimetric formulations, see e.g.\ \cite{Einstein:1916, landau, Rosen:1985gw}. Alternatively, it is reasonable to look for so-called quasi-local quantities associated to 3-volumes $V$ and their boundary 2-surfaces $S$, see e.g.\ \cite{Hawking:1968qt, Penrose:1982wp, Bartnik:1989zz, Brown:1992br}.  In particular, in a recent interesting proposal, Wang and Yau attempt to formulate a reasonable definition of quasi-local energy by considering isometric embeddings in flat spacetime, hence overcoming the problem of background independence \cite{yau1, yau2}.

In this paper, we consider the issue of a background geometry and its seeming necessity for a reasonable definition of quasi-local energy by focusing on Penrose's proposal for quasi-local energy in the context of Kerr-Schild spacetimes.  Kerr-Schild spacetimes, which include the Kerr(-Newman) black hole and gravitational wave solutions, are a simple, yet non-trivial, class of metrics with a natural “background” spacetime to work on. 

The Penrose charge integral \cite{Penrose:1982wp, Penrose:1986ca} (see also Ref.~\cite{todpenrose}) is an energy-momentum construction based on the form of the corresponding charge integral in the weak-field limit.  Penrose showed that in flat spacetime, one can write the energy-momentum in some volume in terms of a surface integral involving twistors. The simplest way of understanding twistors is that they act as potentials for Killing vectors. He then generalises this construction to curved spacetime. However, because twistors do not generally exist in curved spacetime \cite{Lewandowski}, the notion of twistors is replaced by the notion of a 2-surface twistor, which is a projection of the twistor equation transverse to the surface.  The 2-surface twistor equation is generally hard to solve and the Penrose mass has only been calculated for certain special surfaces, such as spherically symmetric surfaces in spherically symmetric spacetimes \cite{Tod:1983waa, Tod:1986gw}. 

We exploit the existence of a natural Minkowski background for Kerr-Schild spacetimes as well as the relation of the Penrose charge integral to the Nester-Witten 2-form \cite{Szabados:1994gw} to modify the original Penrose construction.  In particular, we extend the Penrose construction from the spinor bundle to the bundle of linear frames.  The modified Penrose quasi-local energy is then given by a basis transformation from the induced null tetrad to the Cartesian coordinate basis, which is simply the orthonormal basis of the Minkowski background associated with the Kerr-Schild spacetime. The modified Penrose energy is then equal to the integral of the Nester-Witten 2-form, evaluated on the linear frame bundle using the Cartesian coordinate basis over the surface bounding the volume of interest~\footnote{A brief description of the conventions is provided at the end of the introduction.}
\begin{equation}
 P^{\text{KS}}_{\ba} = \frac{1}{8\pi}\int_S  W_{\ba}[\delta^{a}_{\ba}].
\end{equation}
Using Stokes' theorem, the above expression can be rewritten as a volume integral, which, using properties of the Kerr-Schild ansatz, reduces to
\begin{equation}
  P^{\text{KS}}_{\ba} = \frac{1}{8\pi}\int_V  G_{\ba}{}^{\bb} \Sigma_{\bb},
\end{equation}
i.e.\ a volume integral of the Einstein tensor.  The above charge integral has been proposed before for Kerr-Schild spacetimes \cite{pseudoten,pseudoten1, pseudoten2}.  In particular, it coincides with the Einstein and Landau-Lifshitz pseudotensors.  However, there are also other inequivalent proposals for the Kerr-Schild quasi-local energy \cite{pten3}.  Furthermore, from a physical point of view, considering a Kerr-Schild spacetime as an exact perturbation on Minkowski spacetime, it would make sense that the Einstein tensor represent the gravitational energy-momentum tensor, since the Einstein tensor vanishes for Minkowski spacetime.

In addition, given the close relation between Kerr-Schild and Minkowski spacetimes, one would expect that there is a relation between solutions of the 2-surface twistor equation on a Kerr-Schild background and twistors in Minkowski spacetime.  We investigate also this possibility and find that indeed for the case where the Kerr-Schild null vector is orthogonal to the surface and shear-free,~\footnote{The Goldberg-Sachs theorem \cite{GS} guarantees this to be the case for vacuum Einstein solutions.} there is a one-to-one correspondence between the 2-surface twistors on the Kerr-Schild background and those on Minkowski spacetime.

While the results in this paper are restricted to the setting of Kerr-Schild spacetimes, which have the advantage of admitting a natural background, and it is hard to see how one can generalise the construction in this paper, the result is significant in that it not only emphasises the importance of a background geometry, but also, perhaps more importantly the significance of the bundle on which the quasi-local energy is defined.  The fact that the equivalence principle precludes a tensorial energy-momentum quantity makes pseudotensors, viewed as sections of a particular bundle, the only real viable option. 

We begin in sections \ref{sec:KS} and \ref{sec:penrose} by giving a brief review of Kerr-Schild metrics and the Penrose quasi-local energy, respectively.  In section \ref{sec:modPen}, we describe how the Penrose construction can be modified for the class of Kerr-Schild metrics, leading to a volume integral of the Einstein tensor.  In section \ref{sec:2twistors}, we construct a 1-1 map between 2-surface twistors in Kerr-Schild backgrounds and those in Minkowski spacetime for the special case where the Kerr-Schild null vector is normal to the surface of interest.  We finish in section \ref{sec:ADM} by showing that our proposed modification reduces to the ADM energy in the global case.  Appendix \ref{sec:app} provides a brief review of spinors and the GHP formalism, knowledge of which will be assumed in the main text of the paper. 

\paragraph{Conventions and notation:}
 We work with geometrised units, the metric signature is $-2$, the Riemann (curvature) tensor $R_{abc}^{\phantom{abc}d}$ is defined by
\begin{equation*}
(\nabla_a\nabla_b-\nabla_b\nabla_a)\omega_c=R_{abc}^{\phantom{abc}d}\omega_d
\end{equation*}
for any derivative operator $\nabla_a$ and dual vector field $\omega_d$.
The symmetrisation of a tensor $T_{ab}$ is denoted by 
\begin{equation*}
T_{(ab)}=\frac{1}{2!}(T_{ab}+T_{ba})
\end{equation*}
and the antisymmetrisation by 
\begin{equation*}
T_{[ab]}=\frac{1}{2!}(T_{ab}-T_{ba}),
\end{equation*}
with the corresponding generalisations for tensors and spinors of arbitrary valence. We use the abstract index notation \cite{Penrose:1986vol1ca} and denote discrete indices with bold Latin letters $(\ba,\bb,\ldots)$ or bold Greek letters $(\boldsymbol{\alpha}, \boldsymbol{\beta},\ldots)$ for orthonormal frames, for example a vector basis is denoted by $(\delta_{\ba}^a)$, where $\ba = 0,\dots,3$.  Latin letters $(a,b, \ldots)$ denote tangent space indices and capital Latin letters $(A,B,\ldots)$ denote spinor indices.  Furthermore, in sections \ref{sec:penrose} and \ref{sec:modPen}, we use Greek letters $(\alpha, \beta, \ldots)$ also to denote sections of the twistor bundle.  It should be clear from the context what the Greek letter indices denote.

\section{Kerr-Schild metrics} \label{sec:KS} 
The class of Kerr-Schild metrics can be thought of as exact linear perturbations on flat Minkowski spacetime.~\footnote{See Ref.\ \cite{stephani} for an introduction.}  Working in Cartesian coordinates $(t,x,y,z)$, the Kerr-Schild class is defined by the following line element
\begin{equation} \label{metric:KS}
g_{\ba\bb} = \eta_{\ba\bb}-2 S k_{\ba} k_{\bb}, 
\end{equation}
where $\eta_{\ba\bb}$ is the standard Minkowski line element in Cartesian coordinates and $S$ and $k$ are a scalar function and a null vector defined on the background Minkowski spacetime, respectively.

We generally focus on Kerr-Schild solutions of the Einstein equation where the matter energy-momentum tensor satisfies 
\begin{equation} \label{KS:em}
 T_{ab}\ k^{a}\, k^{b}=0.
\end{equation}
This includes vacuum Kerr-Schild spacetimes, which are sufficient for gravitational energy constructions.

Assuming that equation \eqref{KS:em} is satisfied, we list some important properties of Kerr-Schild metrics \cite{stephani}:
\begin{equation} \label{KS:geoPND}
 k^a \nabla_a k^{b} = 0, \qquad k^{a} k^{c}\; C_{abcd} = \left(D^2 S - R/6 \right) k_{b} k_{d},
\end{equation}
where $D \equiv k^a \partial_a$.  The first equation gives that $k$ is an affinely parametrised geodesic, while the second equation implies that $k$ is a multiple principal null direction of the Weyl tensor $C_{abcd}$, and hence, the spacetime is algebraically special.  Moreover, the Einstein tensor is simply given by
\begin{equation}\label{KS:einstein}
G_{a}^{\phantom{\ua}b}=2\eta_{a c}\partial_{d}\partial_{e} (S k^{e} k^{[b} \eta^{d]c}-Sk^{c}k^{[b}\eta^{d]e}).
\end{equation}

An important example of such a metric is that corresponding to the Kerr-Newman black hole given by
\begin{gather} \label{KS:kerr}
S=\frac{2MR^3-Q^2R^2}{2(R^4+a^2z^2)}, \notag \\[2mm]
(k_0,k_1,k_2,k_3)=(1,\frac{Rx+ay}{R^2+a^2},\frac{Ry-ax}{R^2+a^2},\frac{z}{R})
\end{gather}
with
\begin{equation}
x^2+y^2+z^2=R^2+a^2(1-\frac{z^2}{R^2}).
\end{equation}

\section{Penrose quasi-local energy} \label{sec:penrose}
The Penrose charge integral associated to generic 2-surfaces $S$ of spherical topology is given by~\footnote{The construction is introduced in Ref.\ \cite{Penrose:1982wp} and is thoroughly discussed in chapter 9 of Ref.~\cite{Penrose:1986ca}. For an introduction to 2-surface twistors and the compacted spin-coefficient formalism, see Refs.\ \cite{Penrose:1986vol1ca, Penrose:1986ca}.}
\begin{equation}\label{def:penrose}
A_S[\omega^A,\lambda^A]\equiv \frac{i}{8\pi}\int_S\omega^A\lambda^B R_{ABcd}dx^c\wedge dx^d,
\end{equation} 
where $\omega^A$ and $\lambda^B$ solve the 2-surface twistor equation
\begin{equation} \label{eqn:2twistor}
\eth \omega^1=\sigma \omega^0,\quad \eth' \omega^0=\sigma' \omega^1
\end{equation} 
for the components of the spinor fields $\omega^A$ and $\lambda^A$ with respect to a GHP spin frame $(o^A,\iota^A)$ and
\begin{equation}
 R_{ABcd}=\frac{1}{2}R_{A\pA B A' cd}^{\pA A'}
\end{equation}
with $R_{abcd}$ the Riemann curvature tensor. 

Penrose's charge integral and the corresponding propagation law for the spinor fields can be derived by considering the weak-field limit of general relativity on Minkowski spacetime, as we will now briefly review following Ref.\ \cite{Szabados:2004vb}.  As for global energy constructions in the weak-field approximation and the gravitational mass in Newtonian gravity, the quasi-local energy of the gravitational field is expected to be a surface integral of the curvature associated to an extended, but finite 2-surface $S$. It should be equal to the charge integral of the non-gravitational fields acting as a source for the curvature. Thus, we expect that 
\begin{equation}
\int_VK_aT^{ab}\Sigma_b= \frac{1}{8\pi}\int_S\omega^{AB} R_{ABcd}dx^c\wedge dx^d,
\end{equation}
where $K_a$ is a Killing vector, $T_{ab}$ the (linearised) matter energy-momentum tensor, $\omega^{AB}$ an arbitrary symmetric $(2,0)$-spinor and $\Sigma_b$ the volume 3-form. By means of the Einstein equation, the integrand on the left-hand side equals the exterior derivative of the integrand on the right-hand side iff $\omega^{AB}$ solves the valence $2$ twistor equation and acts as a potential for the Killing vector $K_a$. The above integral $A_S[\omega^A,\lambda^A]$, defined in equation \eqref{def:penrose}, is simply the generalisation of the above flat spacetime construction to curved spacetime with $\omega^{AB} = \omega^{(A} \lambda^{B)}$ and the 2-surface twistor equation \eqref{eqn:2twistor} replacing the twistor equation on Minkowski spacetime as the propagation law. 

The curvature integral $A_S$ can be considered to constitute a bilinear map in $\omega^A$ and $\lambda^B$, and defines by its value the so-called angular momentum twistor $A_{\alpha \beta}$
\begin{equation}
A_{\alpha\beta}\underset{1}{Z}^{\alpha}\underset{2}{Z}^{\beta}\equiv iA_S[\omega^A,\lambda^A]
\end{equation}
for 2-surface twistors $\underset{1}{Z}^{\alpha}=(\omega^A,i\Delta_{A'A}\omega^A)$ and $\underset{2}{Z}^{\alpha}=(\lambda^A,i\Delta_{A'A}\lambda^A)$ with $\Delta_{AA'}$ the 2-dimensional Sen operator \cite{Szabados:2004vb, Szabados:1994fe}.  The 2-dimensional Sen operator $\Delta_a$ is simply the projection of the Levi-Civita connection to the tangent space of the 2-surface $S$,
\begin{equation}
\Delta_a=\Pi_a{}^b\nabla_b,
\end{equation}
where $\Pi_a{}^b$ is the projection operator onto the surface.  More precisely, given an orthonormal set of timelike and spacelike vectors $(t^a, v^a)$, respectively, normal to the surface $S$,
\begin{equation}
 \Pi_a{}^b= \delta^b_a-t_at^b+v_av^b.
\end{equation}

The energy-momentum defined by the Penrose charge integral is an off-diagonal spinor part of the angular momentum twistor $A_{\alpha\beta}$, which can be extracted in terms of the following contraction 
\begin{align}\label{KinTwist:curved}
P^{AA'}\pi_{A}\bar{\pi}_{A'}
&=-A_{\alpha\beta}Z^{\alpha}I^{\beta\gamma}\bar{Z}_{\gamma}\nonumber\\
&=A_S[\epsilon^{AB}\Delta_{BB'}\bar{\omega}^{B'},\omega^{A}]\nonumber\\
&=\frac{1}{4\pi}\int_SW[\pi_A],
\end{align}
where
\begin{equation}\label{KinTwist:W}
W[\pi_A]\equiv i\bar{\pi}_{A'}\nabla_{BB'}\pi_A dx^b\wedge dx^a, \qquad \bar{\pi}_{A'}=i \nabla_{AA'} \omega^A
\end{equation}
is the Nester-Witten 2-form on the bundle of spin frames and $I^{\beta\gamma}$ is the infinity twistor, which ensures that the overall contraction with the angular momentum twistor yields the left-hand side of equation \eqref{KinTwist:curved}.~\footnote{See the beginning of section \ref{sec:modPen} for a discussion of the problems associated with this construction.} The last equality in \eqref{KinTwist:curved} follows from the 2-surface twistor equation \cite{Szabados:1994gw}. The Nester-Witten 2-form can be extended to the bundle of linear frames by a vierbein $\theta_{\ba}^{\pa \boldsymbol{\alpha}}$ and the transformation matrix $A_{\boldsymbol{\alpha}}^{\pa\ba}$ relating the induced null basis to the Minkowski basis according to
\begin{equation}
\begin{xy}
\xymatrix{
	\epsilon_{\bA}^{\pA A} \ar[r]^{\otimes}   &  n_{\ba}^a \ar[r]^{A_{\boldsymbol{\alpha}}^{\pa \ba}}   &  \theta_{\boldsymbol{\alpha}}^a \ar[r]^{\theta^{\pa \boldsymbol{\alpha}}_{\ba}  }&\delta_{\ba}^a
}
\end{xy},
\end{equation}
where $(\epsilon_{\bA}^{\pA A})$ is a spin frame, $(n_{\ba}^a)$ the null basis obtained by the tensor product of $(\epsilon_{\bA}^{\pA A})$ with itself, $(\theta_{\boldsymbol{\alpha}}^a)$ the canonical Minkowski basis given by a constant linear combination of the induced null basis and $(\delta_{\ba}^a)$ the Cartesian coordinate basis.~\footnote{We denote the discrete index of the induced Minkowski basis by Greek letters in order to distinguish the membership of the different indices to the different bases.}  Thus, considering the spin frame $\epsilon_{A}^{\pA\bA}$ with $\epsilon_{A}^{\pA0}=\pi_A$, the $00$-component of the energy-momentum is given by the Nester-Witten integral
\begin{align*}
P^{00}&=P^{AA'}\pi_{A}\bar{\pi}_{A'}\\
&=\frac{1}{8\pi} A^{\boldsymbol{\alpha}}_{\pa 3}\int_S \theta^{\ba}_{\pa \boldsymbol{\alpha}}W_{\ba}[\delta^a_{\ba}],
\end{align*}
where
\begin{equation}\label{NW:linear}
W_{\ba}[\delta_{\ba}^a]=\frac{1}{2}\epsilon_{\ba\bb\bc\boldsymbol{d}}dx^{\bb}\wedge\omega^{\bc\boldsymbol{d}}
\end{equation}
is the extension of the Nester-Witten 2-form $W[\pi_A]$ to the bundle of linear frames with $\omega^{\bc\boldsymbol{d}}=\omega_{\bb}^{\pb \bc\boldsymbol{d}}dx^{\bb}$ and $\omega_{\ba\phantom{\nu}\boldsymbol{\nu}}^{\pa\boldsymbol{\mu}}$ being the spin connections defined according to $\nabla_{\ba}\theta_{\boldsymbol{\nu}}^a=\omega_{\ba\phantom{\nu}\boldsymbol{\nu}}^{\pa\boldsymbol{\mu}}\theta_{\boldsymbol{\mu}}^a$. We denote the inverse transformations of $A_{\boldsymbol{\alpha}}^{\pa\ba}$ and $\theta_{\ba}^{\pa \boldsymbol{\alpha}}$ by $A^{\boldsymbol{\alpha}}_{\pa\ba}$ and $\theta^{\ba}_{\pa \boldsymbol{\alpha}}$, respectively.

\section{Modifying the Penrose energy}\label{sec:modPen}
The contraction $P^{AA'}\pi_A\bar{\pi}_{A'}$ is expected to be constant on the surface $S$ as can be seen from its definition in terms of the original Penrose charge integral, \textit{cf.} equation \eqref{KinTwist:curved}. This constancy as well as other issues such as the existence of an infinity twistor and hence, the well-definedness of the overall contraction $P^{AA'}\pi_{A}\bar{\pi}_{A'}
=-A_{\alpha\beta}Z^{\alpha}I^{\beta\gamma}\bar{Z}_{\gamma}\nonumber$ is an outstanding problem. In fact, Helfer \cite{Helfer:1986gw} claims that no such infinity twistor exists on curved spacetimes and the particular combination $P^{AA'}\pi_{A}\bar{\pi}_{A'}$ may neither be constant on $S$ nor extractable from $A_{\alpha\beta}$ at all. Therefore, he proposes to consider the Nester-Witten 2-form integral directly as the defining value of the contraction 
\begin{equation}
 P^{AA'}\pi_{A}\bar{\pi}_{A'} \equiv \frac{1}{4\pi}\int_SW[\pi_A]. 
\end{equation}
For the following discussion, we assume that the original Penrose construction is well-defined for Kerr-Schild spacetimes or, alternatively, adopt the suggestion of Helfer and consider the Nester-Witten 2-form integral evaluated on the bundle of spin frames as the defining value of the contraction $P^{AA'}\pi_{A}\bar{\pi}_{A'}$. We discuss 2-surface twistors on Kerr-Schild spacetimes in section \ref{sec:2twistors}.

By the assumed constancy, we can write $P^{00}$ as
\begin{align}\label{P:integral}
P^{00}&=\frac{1}{\text{area}(S)}\int_S dS \text{ }P^{AA'}n_a^0\nonumber\\
&=\frac{1}{8\pi} A^{\boldsymbol{\alpha}}_{\pa 3}\int_S \theta^{\ba}_{\pa \boldsymbol{\alpha}}W_{\ba}[\delta^a_{\ba}]
\end{align}
and interpret the integrand as the 0-component of the energy-momentum surface density with respect to the induced null tetrad with $n_a^0=\pi_A\bar{\pi}_{A'}$. Therefore, the quasi-local energy-momentum with respect to the Cartesian coordinate basis is given by 
\begin{align}\label{KS:integral}
P^{\text{KS}}_{\ba}&\equiv \frac{1}{\text{area}(S)}\int_SdS\text{ } g_{\ba\bb}P^{AA'}\delta_a^{\bb}.
\end{align}
This integral can be related to the above $P^{00}$ as follows. The 2-surface twistor equation has generically exactly four linearly independent solutions for surfaces of spherical topology \cite{baston} (see, however, Ref.\ \cite{jeff}). These four solutions define four spin frames with $\underset{i}{\epsilon}\text{}_A^{\pA 0}=\underset{i}{\pi}\text{}_A $ and four induced null frames $\underset{i}{n}\text{}_a^{\ba} $. Thus, we have four integrals of the type \eqref{P:integral}, denoted by 
\begin{align}\label{em:charges}
\underset{i}{P}^{0}&\equiv\frac{1}{\text{area}(S)}\int_S dS\text{ } P^{AA'}\underset{i}{n}\text{}_a^{0}\nonumber\\
&=\frac{1}{8\pi} A^{\boldsymbol{\alpha}}_{\pa 3}\int_S \underset{i}{\theta}\text{}_{\pa \boldsymbol{\alpha}}^{\ba}W_{\ba}[\delta^a_{\ba}]
\end{align}
and obtain a new basis by combining the 0-th basis vectors of these null frames $(\underset{i}{n}\text{}_a^{0} )_{i=0,\dots, 3}$. These vectors are indeed linearly independent over $\mathbb{R}$, since two of them are linearly independent over $\mathbb{C}$. The coordinate basis can be rewritten in terms of this new basis via a basis transformation
\begin{align}\label{basis:transf}
\delta_a^{\ba}&=\sum_i D_i^{\pb \ba}\underset{i}{n}\text{}_a^0,
\end{align}
which provides a relation between $P^{\text{KS}}_{\ba}$ and the integrals $\underset{i}{P}^{0}$. First, note that if we expand the above basis transformation in terms of the induced Minkowski tetrads and the corresponding vierbein followed by the transformation to the coordinate basis, we obtain 
\begin{align*}
\delta_a^{\ba}&=\sum_i D_i^{\pa\ba}\underset{i}{n}\text{}_a^0\\
&=\sum_i D_i^{\pa\ba}A^{\boldsymbol{\alpha} }_{\pa 3}g_{\bb\bc}\underset{i}{\theta}\text{}^{\bc}_{\pc\boldsymbol{\alpha}}\delta_a^{\bb}.
\end{align*}
A comparison of the coefficients of the above expansions yields
\begin{align}\label{kintwist:transf}
\sum_i D_i^{\pa\ba}A^{\boldsymbol{\alpha} }_{\pa 3}\underset{i}{\theta}\text{}^{\bb}_{\pc\boldsymbol{\alpha}}&=g^{\ba\bb}.
\end{align}
Therefore, from equations \eqref{KS:integral}-\eqref{kintwist:transf} we obtain
\begin{align}\label{KS:energy-momentum}
P^{\text{KS}}_{\ba}&=\frac{1}{8\pi}\int_Sg_{\ba\bc}\sum_i D_i^{\pa\bc}A^{\boldsymbol{\alpha}}_{\pa 3} \underset{i}{\theta}\text{}^{\bb}_{\pa\boldsymbol{\alpha}}W_{\bb}[\delta_{\bb}^a]\nonumber\\
&=\frac{1}{8\pi}\int_S W_{\ba}[\delta_{\ba}^a].
\end{align}
Stokes' theorem allows us to rewrite the above surface integral in terms of a volume integral
\begin{equation} \label{KS:vol}
P^{\text{KS}}_{\ba}=\frac{1}{8\pi}\int_V dW_{\ba}[\delta_{\ba}^a]
\end{equation}
and as we find below this turns out to be a particularly interesting quantity. 

Note that the Sparling equation \cite{sparling1, sparling2, mason} gives a beautiful relation between the exterior derivative of the Nester-Witten 2-form and the Einstein tensor
\begin{equation} \label{sparling}
 dW_{\boldsymbol{\alpha}}[\theta^a_{\boldsymbol{\alpha}}] = - G_{\boldsymbol{\alpha}}{}^{\boldsymbol{\beta}} \Sigma_{\boldsymbol{\beta}} - S_{\boldsymbol{\alpha}}[\theta^a_{\boldsymbol{\alpha}}],
\end{equation}
where $S_{\boldsymbol{\alpha}}[\theta^a_{\boldsymbol{\alpha}}]$ is the Sparling 3-form and
\begin{equation}
 \theta^{\boldsymbol{\alpha}} \wedge \theta^{\boldsymbol{\beta}} \wedge \theta^{\boldsymbol{\gamma}} = \epsilon^{\boldsymbol{\alpha} \boldsymbol{\beta} \boldsymbol{\gamma} \boldsymbol{\delta}} \Sigma_{\boldsymbol{\delta}}.
\end{equation}
However, the above equation \eqref{sparling} is clearly only valid on the bundle of orthonormal frames.~\footnote{In Ref.\ \cite{mason}, the Sparling equation has been used directly on the bundle of linear frames leading them to conclude that the Sparling form vanishes for Kerr-Schild metrics.  When computed on the bundle of orthonormal frames, it is clear that the Sparling form is non-zero.}

In what follows, we calculate the Nester-Witten 2-form and its exterior derivative for the Kerr-Schild class on the bundle of linear frames.  We express the vierbein relating the orthonormal frame $(\theta_{\boldsymbol{\alpha}}^a)$ induced by the 2-surface twistor equation to the coordinate basis $(\delta_{\ba}^a)$, such that $g_{\ba\bb}=\theta_{\ba}^{\pa \boldsymbol{\alpha}}\theta_{\bb}^{\pa \boldsymbol{\beta}}\eta_{\boldsymbol{\alpha}\boldsymbol{\beta}}$, as (\textit{cf.} Ref.\ \cite{ee})
\begin{equation}\label{KS:vierbein}
\theta_{\ba}^{\pa \boldsymbol{\alpha}}=e_{\ba}^{\pa \boldsymbol{\alpha}}-Sk_{\ba} k^{\boldsymbol{\alpha}},
\end{equation}
where $k^{\boldsymbol{\alpha}}= e_{\ba}^{\pa \boldsymbol{\alpha}} k^{\ba}=\theta_{\ba}^{\pa \boldsymbol{\alpha}} k^{\ba}$
and $e_{\ba}^{\pa \boldsymbol{\alpha}}$ is the vierbein connecting the induced orthonormal frame to the background Minkowski basis. Expanding the torsion-free, metric compatible covariant derivative of an arbitrary vector with respect to the orthonormal and the coordinate basis yields the following relation between the spin connections and the vierbein $\theta^{\pa\boldsymbol{\alpha}}_{\ba}$
\begin{equation}\label{KS:connection}
\omega_{\ba}^{\pa\boldsymbol{\alpha}\boldsymbol{\beta}}=\theta^{\bb[\boldsymbol{\alpha}}(2\partial_{[\ba}\theta_{\bb]}^{\pb\boldsymbol{\beta}]}+\theta^{\boldsymbol{d}|\boldsymbol{\beta}]}\theta_{\ba\boldsymbol{\gamma}}\partial_{\boldsymbol{d}}\theta_{\bb}^{\pd\boldsymbol{\gamma}}).
\end{equation}
The evaluation of equation \eqref{KS:connection} for the vierbein given by \eqref{KS:vierbein} leads to
\begin{equation}\label{KS:connection_evaluated}
\omega_{\ba}^{\pa\boldsymbol{\alpha}\boldsymbol{\beta}}=2e^{\bb[\boldsymbol{\alpha}}\partial_{\bb}(Sk_{\ba}k^{\boldsymbol{\beta}]})
\end{equation}
for the spin connection of Kerr-Schild metrics. This can be used to evaluate the Nester-Witten 2-form given by \eqref{NW:linear}. The final expression is 
\begin{equation}\label{KS:NW}
W_{\ba}[\delta_{\ba}^a]=\frac{1}{4}g_{\ba\boldsymbol{e}}\partial_{\bb}(g^{\boldsymbol{e}\bc}g^{\bb\boldsymbol{d}}-g^{\boldsymbol{e}\boldsymbol{d}}g^{\bb\bc})\Sigma_{\bc\boldsymbol{d}},
\end{equation}
where $\Sigma_{\ba\bb}=\frac{1}{2}\epsilon_{\ba\bb\bc\boldsymbol{d}}dx^{\bc}\wedge dx^{\boldsymbol{d}}$. The exterior derivative of \eqref{KS:NW} turns out to be the Einstein tensor (see equation \eqref{KS:einstein}) contracted with the volume 3-form $\Sigma_{\ba}=\frac{1}{6}\epsilon_{\ba\bb\bc\boldsymbol{d}}dx^{\bb}\wedge dx^{\bc}\wedge dx^{\boldsymbol{d}}$ for Kerr-Schild solutions of the Einstein equation where the matter energy-momentum tensor satisfies $T_{a b}k^{a}k^{b}=0$, i.e.\
\begin{equation}\label{KS:dNW}
dW_{\ba}[\delta_{\ba}^a]=G_{\ba}^{\pa \bb}\Sigma_{\bb}.
\end{equation}
Substituting the above equation into equation \eqref{KS:vol} gives that the quasi-local energy-momentum of Kerr-Schild spacetimes given by \eqref{KS:energy-momentum} is the Einstein tensor integral
\begin{equation}\label{KS:emEinstein}
P_{\ba}^{\text{KS}}=\frac{1}{8\pi}\int_V G_{\ba}^{\pb\bb}\Sigma_{\bb}.
\end{equation}
While the original Penrose charge integral was only defined for the generic surfaces of spherical topology allowing for a 4 complex-dimensional 2-surface twistor space, the Einstein tensor integral is defined for arbitrary volumes in Kerr-Schild spacetimes. Thus, our construction may be evaluated for arbitrary surfaces.

\section{2-surface twistors on Kerr-Schild spacetimes}\label{sec:2twistors}
In this section, we would like to address some questions associated with the 2-surface twistor space for Kerr-Schild spacetimes.

For simplicity, we consider 2-surfaces $S$ in Kerr-Schild spacetimes for which the geodesic null vector $k$ is a normal vector.~\footnote{The more general setting is too unwieldy for now.  However, it would be very interesting to generalise the results of this section to arbitrary 2-surfaces.} In this setting, the spin frame $(\kappa^A,\iota^A)$ with $k^a=\kappa^A\bar{\kappa}^{A'}$ is a GHP frame on $S$ and induces a null tetrad of the form $(k^a,n^a,m^a,\bar{m}^a)$. Thus, the Kerr-Schild line element can be expressed as 
\begin{align*}
g_{ab}&= \eta_{ab}-2Sk_ak_b\\
&=2k_{(a}n_{b)}-2m_{(a}\bar{m}_{b)}.
\end{align*}
Rearranging the above equation then gives a null tetrad for the Minkowski background 
\begin{align}
\eta_{ab}&= g_{ab}+2Sk_ak_b\nonumber\\
&=2k_{(a}(n_{b)}+Sk_{b)})-2m_{(a}\bar{m}_{b)}.
\end{align}
Hence, the tetrad $(k^a,n^a-Sk^a,m^a,\bar{m}^a)$ is a null tetrad with respect to the Minkowski background with $k$ normal to $S$. The spin coefficients appearing in the 2-surface twistor equation with respect to the GHP frame $(\kappa^A,\iota^A)$ are 
\begin{equation}
\sigma = m^a\delta k_a,\quad \sigma'=\bar{m}^a\delta' n_a,\quad
\beta =\frac{1}{2}(n^a\delta k_a+m^a\delta\bar{m}_a),\quad \beta' =\frac{1}{2}(k^a\delta' n_a+\bar{m}^a\delta' m_a),
\end{equation}
where $\delta=m^a \nabla_a$ and $\delta'=\bar{m}^a \nabla_a$, see e.g.\ \cite{Penrose:1986ca, Penrose:1986vol1ca}. Due to the geodesic and null properties of $k$, the only change in the above spin coefficients due to a change of basis from $(k^a,n^a,m^a,\bar{m}^a)$ to $(k^a,n^a-Sk^a,m^a,\bar{m}^a)$ occurs in $\sigma'$, which transforms as 
\begin{equation*}
\sigma'\rightarrow \sigma'+S \bar{\sigma}.
\end{equation*}
From section \ref{sec:KS} and equation \eqref{KS:geoPND} in particular, we know that \textit{aligned}~\footnote{See equation \eqref{KS:em}.} Kerr-Schild spacetimes are algebraically special.  Therefore, the Goldberg-Sachs theorem guarantees that for vacuum Kerr-Schild spacetimes, the null vector $k$ is shear-free and hence, $\sigma=0$.  More generally, if the congruence of $k$ is indeed shear-free, the 2-surface twistor equation for the components of the spinor solution with respect to $(\kappa^A,\iota^A)$ are the same as the 2-surface twistor equation for the components on the Minkowski background with respect to the spin frame with induced null tetrad $(k^a,n^a-Sk^a,m^a,\bar{m}^a)$. Since the latter are simply tangential projections of the twistor equation on the Minkowski background, we find a one-to-one correspondence between the 2-surface twistors for surfaces with normal $k$ and the background Minkowski twistors for Kerr-Schild spacetimes.  For the case where the congruence is shearing, we know that the solution is of Petrov type N, see section 32.4.2 of Ref.\ \cite{stephani}.

\section{Relation to pseudotensors and ADM energy} \label{sec:ADM}
The notion that the Einstein tensor acts as an energy-momentum density for Kerr-Schild spacetimes is supported by other energy-momentum constructions. 

The energy-momentum pseudotensors of Einstein, Landau and Lifshitz, Tolman, Papapetrou, and Weinberg are proportional to the Einstein tensor with respect to the Cartesian coordinates of Kerr-Schild spacetimes \cite{pseudoten2}.

Furthermore, the quantity $P^{\text{KS}}_0$ reproduces the ADM energy for asymptotically flat Kerr-Schild spacetimes in the appropriate limit. The ADM energy is the numerical value of the Hamiltonian surface integral of linearised gravity \cite{Arnowitt:1962hi} and is given by
\begin{equation}\label{ADM}
E_{\text{ADM}}= \frac{1}{16\pi}\lim_{r\rightarrow \infty} \int_{S^2_r} d^2S\ n_i (\partial_jh_{ij}-\partial_ih_{jj}),
\end{equation}
where $h_{ij}$ is the linear perturbation from the flat (spatial) metric in terms of asymptotically flat Euclidean coordinates, the summation over the spatial indices $i$ and $j$ runs from $1$ to $3$ and the vector $n^{i}$ is the unit outward normal of the 2-sphere $S^2_r$ with asymptotically Euclidean radius $r$. Considering asymptotically flat Kerr-Schild spacetimes, where the components of the Kerr-Schild deviation in terms of the Cartesian coordinate basis fall off sufficiently fast at large radii $r=\sqrt{x^ix^i}$ and a potential time dependence at spatial infinity does not contribute to the final integral, equations \eqref{KS:einstein} and \eqref{KS:emEinstein} lead to
\begin{equation}\label{KS:AFe}
P^{\text{KS}}_0= -\frac{1}{8\pi}\int_{S} d^2S\ n_i (\partial_j(Sk_ik_j)-\partial_i(Sk_jk_j)).
\end{equation}
The deviation from the induced spatial metric of the asymptotically flat Kerr-Schild spacetime is $h_{ij}=-2Sk_ik_j$, thus from equations \eqref{ADM} and \eqref{KS:AFe} we find
\begin{equation}
P^{\text{KS}}_0=E_{\text{ADM}}|_{h_{ij}=-2Sk_ik_j}.
\end{equation}
Therefore, the Einstein tensor integral satisfies the correct large sphere behaviour at spatial infinity for asymptotically flat Kerr-Schild spacetimes by reproducing the ADM energy.

\section*{Acknowledgements}

M.G.\ is partially supported by grant no.\ 615203 from the European Research Council under the FP7.

\appendix

\section{Spinors and the Geroch-Held-Penrose (GHP) formalism} \label{sec:app}

Spinors are defined as sections of the spinor bundle whose fibre at a point is a 2-dimensional complex vector space, called the spin space.  There exists a unique inner product on this space given by
\begin{equation}
 \{\kappa, \omega\} = \epsilon_{AB} \kappa^A \omega^B,
\end{equation}
where $\epsilon_{AB}$ is an antisymmetric spinor such that
\begin{equation}
 \epsilon_{AC} \epsilon^{BC} = \delta_A{}^B.
\end{equation}
Equivalently, $\epsilon_{AB}$ and $\epsilon^{AB}$ can be used to lower and raise spinor indices, respectively.  Thus,\footnote{This convention for the product of spinors is known as the ``North East-South West'' convention.}
\begin{equation}
 \kappa_A = \kappa^B \epsilon_{BA}, \qquad \kappa^A = \epsilon^{AB} \kappa_{B}.
\end{equation}
We may choose a GHP \cite{NP,GHP} spin frame $(o^A,\iota^A)$, i.e.\ a basis with the spinors normalised such that
\begin{equation} \label{app:oi}
 o_A \iota^A = 1.
\end{equation}

The spinor map, or the isomorphism between the proper, orthochronous Lorentz group SO$^+$(1,3) and SL(2,$\mathcal{C}$), allows one to write vectors as bilinears of spinors.  Hence, we identify tangent bundle indices with spinor indices
\begin{equation}
a=AA', \ldots.
\end{equation}
The metric
\begin{equation} \label{app:metric}
 g_{ab} = \epsilon_{AB} \epsilon_{A' B'}
\end{equation}
and we construct a complex null frame $(\ell^a, n^a, m^a, \bar{m}^a)$ via the following identification
\begin{equation}
 \ell^a = o^A \bar{o}^{A'}, \qquad n^a = \iota^A \bar{\iota}^{A'}, \qquad m^a = o^A\bar{\iota}^{A'}, \qquad \bar{m}^a = \iota^A \bar{o}^{A'},
\end{equation}
where, for example, $\bar{o}^{A'}$ is the complex conjugate of $o^A$.  Equations \eqref{app:oi} and \eqref{app:metric} then imply that the only non-trivial inner products between the null frame vectors are
\begin{equation}
\ell^a n_a=1,\quad m^a\bar{m}_a=-1
\end{equation}
and that
\begin{equation}
g_{ab} = \ell_a n_b + n_a \ell_b - m_a \bar{m}_b - \bar{m}_a m_b.
\end{equation}
Using such a Newman-Penrose \cite{NP} or GHP \cite{GHP} frame, one can construct scalars by taking the components of tensors along the null frame directions.  For example, for a covector $V_{a}$,
\begin{equation}
 V_0 \equiv \ell^a V_a, \qquad  V_1 \equiv n^a V_a, \qquad V_m \equiv m^a V_a, \qquad V_{\bar{m}} \equiv \bar{m}^a V_a.
\end{equation}
Taking the components of the covariant derivatives of the frame vectors along the null frame gives twelve independent complex spin coefficients.  For example, for some components of the derivative of $\ell_a$, we have
\begin{equation} \label{app:kappa}
 \kappa = m^b \ell^a \nabla_a \ell_b, \qquad \sigma = m^b m^a \nabla_a \ell_b, \qquad \rho = m^b \bar{m}^a \nabla_a \ell_b, \qquad \tau = m^b n^a \nabla_a \ell_b.
\end{equation}
Physically, the above scalars parametrise the geodesity, shear, expansion and twist, respectively, of the null congruence of $\ell^a.$  Defining a prime operation so that\footnote{Equivalently, $$ (o^A)' = i\, \iota^A, \qquad (\iota^A)' = i\, o^A.$$}
\begin{equation}
 (\ell^a)' = n^a, \quad  (n^a)'=\ell^a, \quad (m^a)' = \bar{m}^a, \quad (\bar{m}^a)' = m^a,
\end{equation}
$\kappa'$, $\sigma'$, $\rho'$ and $\tau'$ parametrise the  geodesity, shear, expansion and twist, respectively, of the null congruence of $n^a.$  

All the eight spin coefficients defined above are GHP scalars in the following sense: a GHP scalar $\eta$ of spin weight $\{p,q\}$ is defined as a scalar such that under the following transformation  
\begin{equation}
 o^A \rightarrow \lambda\, o^A, \qquad \iota^A \rightarrow \lambda^{-1}\, \iota^A,
\end{equation}
for some arbitrary complex scalar $\lambda$, 
\begin{equation}
 \eta \rightarrow \lambda^p \bar{\lambda}^q\ \eta.
\end{equation}
For example, $\kappa$ is a GHP scalar of weight $\{3,1\}$, as can be easily verified from its definition in equation \eqref{app:kappa}.

Of the twelve independent complex spin coefficients, the remaining four
\begin{equation}
 \beta = \frac{1}{2} (n^a m^b \nabla_{b} \ell_a + m^a m^b \nabla_b \bar{m}_a), \quad \epsilon = \frac{1}{2} ( n^a \ell^b\nabla_{b} \ell_a + m^a \ell^b\nabla_b \bar{m}_a),
\end{equation}
as well as $\beta'$ and $\epsilon'$ are not GHP scalars.  These spin coefficients can be used to define ``GHP covariant'' derivative operators acting on a GHP scalar $\eta$ of spin weight $\{p,q\}$ as follows:
\begin{align}
 \tho \eta &= (\ell^a \nabla_a - p \epsilon - q \bar{\epsilon})\, \eta, \qquad \tho' \eta= (n^a \nabla_a + p \epsilon' + q \bar{\epsilon}')\, \eta, \notag \\
 \eth \eta &= (m^a \nabla_a - p \beta + q \bar{\beta}')\, \eta, \quad \eth'\eta = (\bar{m}^a \nabla_a + p \beta' - q \bar{\beta})\, \eta. 
\end{align}
The spin weights of the derivative operators are
\begin{equation}
 \tho:\ \{1,1\}, \qquad \tho':\ \{-1,-1\}, \qquad \eth:\ \{1,-1\}, \qquad \eth':\ \{-1,1\}.
\end{equation}

\bibliographystyle{utphys}
\bibliography{penrose}

\end{document}